\documentclass[conference,10pt]{IEEEtran}
\usepackage{epsfig,rotating,setspace,latexsym,amsmath,epsf,amssymb,amsfonts,bm,subfigure,epstopdf}
\usepackage{cite,authblk}
\usepackage{bbm}
\usepackage{mathtools,amsthm}
\usepackage{algorithm}
\usepackage[noend]{algpseudocode}

\algrenewcommand\algorithmicforall{\textbf{foreach}}
\algrenewcommand\algorithmicindent{.8em}
\usepackage{comment}
\usepackage{color}
\usepackage{amsmath}
\DeclareMathOperator*{\argmax}{arg\,max}
\DeclareMathOperator*{\argmin}{arg\,min}
\usepackage{mathtools}

\newtheorem{theorem}{Theorem}

\newenvironment{Proof}[1]{\medskip\par\noindent{\bf Proof:\,}\,#1}{{\mbox{\,$\blacksquare$}\par}}
\IEEEoverridecommandlockouts
\allowdisplaybreaks

\begin{document}

\title{Game Theoretic Analysis of an Adversarial \\ Status Updating System}

\author{Subhankar Banerjee \qquad Sennur Ulukus\\
\normalsize Department of Electrical and Computer Engineering\\
\normalsize University of Maryland, College Park, MD 20742\\
\normalsize  \emph{sbanerje@umd.edu} \qquad \emph{ulukus@umd.edu}}	

\maketitle

\begin{abstract}
We investigate the game theoretic equilibrium points of a status updating system with an adversary that jams the updates in the downlink. We consider the system models with and without diversity. The adversary can jam up to $\alpha$ proportion of the entire communication window. In the model without diversity, in each time slot, the base station schedules a user from $N$ users according to a stationary distribution. The adversary blocks (jams) $\alpha T$ time slots of its choosing out of the total $T$ time slots. For this system, we show that a Nash equilibrium does not exist, however, a Stackelberg equilibrium exists when the scheduling algorithm of the base station acts as the leader and the adversary acts as the follower. In the model with diversity, in each time slot, the base station schedules a user from $N$ users and chooses a sub-carrier from $N_{sub}$ sub-carriers to transmit update packets to the scheduled user according to a stationary distribution. The adversary blocks $\alpha T$ time slots of its choosing out of $T$ time slots at the sub-carriers of its choosing. For this system, we show that a Nash equilibrium exists and identify the Nash equilibrium. 
\end{abstract}

\section{Introduction}
Due to the advances in modern applications such as virtual reality, augmented reality, robotics, mission critical control and various other 5G technologies, the \emph{quality of user experience} has become important, perhaps even more important than \emph{quality of service}, which is typically measured in terms of delay, throughput and bit error. A recently introduced quality of user experience metric is \emph{freshness of information} which is measured in terms of the \emph{age of information} \cite{kaul2012real}. In a slotted system, the age of information of a user is $t-\tau$, where $t$ is the current time slot and $\tau$ is the last time slot when the user has received an update packet. Maintaining information freshness at a user is different than maintaining low delay or high throughput at the user; fresh information delivery is achieved by delivering sufficiently frequent packets at sufficiently low delays, which requires operating the system at a novel interior throughput-delay point; see recent surveys \cite{kosta2017age, SunSurvey, YatesSurvey}. 

A vast amount of literature is available now on the analysis and optimization of the age of information metric ranging from various queueing regimes to energy harvesting systems, wireless networks, remote estimation, gossip networks, caching systems, source coding problems, and so on, see e.g., \cite{Najm17, Soysal19, Ayan19, Yates20_moments, Farazi18, Wu18, Baknina18, Leng19, Arafa19TwoHop, gu2020twohop, Arafa20, wireless-ephremides, kadota18, Buyukates19_multihop, hsu18, Buyukates19_hier, Elmagid20, Ceran18, Buyukates20_stragglers, zou19,  yang20, ozfatura20, buyukates-fl, liu18, wang19_counting, bastopcu20_google, sun17_remote, chakravorty20, Bastopcu19_distortion, vaze21jsac, Bastopcu20_soft_updates, Yates_Soljanin_source_coding, mayekar20, Bastopcu20_selective, Yates17sqrt, bastopcu2020LimitedCache, Eryilmaz21, Kaswan-isit2021, Yates21gossip, baturalp21comm_struc, bastopcu21gossip, kam20}. Most of the existing literature considers communication systems without any adversaries. The recent works that are most closely related to our work here are \cite{banerjee2020fundamental, bhattacharjee2020competitive, garnaev2019maintaining, nguyen2017impact, xiao2018dynamic, banarjee-infocom2022}. Specifically, \cite{banerjee2020fundamental, bhattacharjee2020competitive} use an adversary in their communication system to model non-staionarity of a wireless communication channel. The adversary blocks a communication channel by completely eliminating the transmitted update packet, increasing the age. Similar to \cite{banerjee2020fundamental, bhattacharjee2020competitive}, we consider an adversary which blocks a communication link by completely eliminating the update packets. However, different than \cite{banerjee2020fundamental, bhattacharjee2020competitive}, we consider a power constrained adversary; in addition, while \cite{banerjee2020fundamental, bhattacharjee2020competitive} consider competitive ratio, we consider the absolute age performance. \cite{garnaev2019maintaining, nguyen2017impact} consider an adversary that acts as an interferer which decreases the signal-to-noise-ratio of the communication link, decreasing the data rate, ultimately increasing the age. \cite{xiao2018dynamic} considers an adversary which blocks a communication channel for a time duration in continuous time, which results in higher age for the system, by disabling the transmissions for some time after the reception of the last update packet. Unlike \cite{garnaev2019maintaining, nguyen2017impact, xiao2018dynamic}, our considered adversary increases the age by directly jamming (blocking) the channel in time slots of its choosing in a time slotted system. In this work, we extend the model in \cite{banarjee-infocom2022} to a game theoretic setting.

In this paper, the adversary can block any $\alpha T$ time slots over a time horizon of $T$ slots where $\alpha < 1$. First, we investigate the system, where at every time slots, the base station schedules a user from $N$ users and the adversary blocks any one of the $N$ users for any $\alpha T$ time slots. In this case, we do not have any diversity present in the system. For this system, we show that the Nash equilibrium does not exist. However, a Stackelberg equilibrium exists, and we find the Stackelberg equilibrium point. Next, we investigate the system with diversity by introducing diversity in frequency domain via sub-carriers \cite{bahai2004multi}. In this system, at every time slot, the base station schedules a user from $N$ users and chooses a sub-carrier among available $N_{sub}$ sub-carriers. The adversary blocks any one of the $N_{sub}$ sub-carriers for any $\alpha T$ time slots. For this case, we show that the Nash equilibrium exists, and we find the Nash equilibrium point.

\section{System Models and Problem Formulation}\label{sec:2}
We consider two different settings for a wireless communication network. In the first setting, there is no diversity, and in the second setting, we incorporate diversity in the frequency domain by introducing multiple sub-carriers.

\subsection{Communication Network Model without Diversity}\label{subsec:A}
We consider a communication system, where at each time slot, a base station schedules a user from $N$ users following a stationary distribution. The base station transmits an update packet to the scheduled user at each time slot. The adversary present in the system can block a communication channel between a user and the base station. By blocking a communication channel, the adversary completely eliminates the update packet. The adversary only knows the scheduling algorithm used by the base station, and if the base station uses a randomized algorithm, then the adversary knows the probability distribution used by the base station, but does not know specific realization at each time. The adversary can block only one communication channel at a given time slot, and in total it can block $\alpha T$ time slots over the time horizon of $T$ time slots, where $0<\alpha<1$. Let $p_{i}$ be the probability with which the base station schedules user $i$. Thus, the average time needed for the user $i$ to get scheduled is $\frac{1}{p_{i}}$. Hence, we assume that the time horizon $T\gg \frac{1}{p_{i}}$ for all $i$. In addition, $T$ is large enough such that for a given $\alpha$, $T(1-\alpha)$ is also large.
 
Let $\bm{p}=[p_1 ~ p_2 ~ \cdots ~ p_N]$ be the probability mass function with which the base station schedules users at every time slot. Let $\sigma_{i}(t)$ denote the action of the adversary for the communication channel between user $i$ and the base station at time $t$. Here $\sigma_{i}(t) = 0$ means that the adversary blocks user $i$ at time $t$, and $\sigma_{i}(t)=1$ means that the adversary does not block user $i$ at time $t$. Thus, the action of the adversary against user $i$ a sequence of ones and zeros. We denote this sequence with $\sigma_{i}$, and we call this sequence a \emph{blocking sequence} for user $i$. We use $\sigma$ to denote the blocking matrix whose $i$th row is the blocking sequence $\sigma_{i}$, and thus, $(i,t)$th entry $\sigma_i(t)$ is the state of the communication channel between user $i$ and the base station at time $t$. Thus, a feasible $\sigma$ should satisfy $\sum_{i=1}^{N}\sum_{t=1}^{T} \sigma_i(t) \leq \alpha T$. We create a set $\Sigma$ of feasible $\sigma$. 

The average age for a user scheduling algorithm, i.e, a probability distribution $\bm{p}$  and a blocking matrix $\sigma$ is 
\begin{align} \label{objective}
\Delta^{\bm{p},\sigma} = \limsup_{T \to \infty} \frac{1}{T} \sum_{t=1}^{T} \frac{1}{N} \left(\sum_{i=1}^{N} \mathbb{E}\left[v_{i}(t)\right]\right)
\end{align}
where $v_{i}(t)$ is the age of user $i$ at time $t$. \cite{banarjee-infocom2022} shows that the age of the communication system for this system model increases linearly with $T$, thus, $\Delta^{\bm{p},\sigma}$ becomes infinity. For this reason, we analyze this system model for \emph{large and finite} $T$ and define the average age for this system model as
\begin{align} \label{objective}
\Delta^{\bm{p},\sigma} =  \frac{1}{T} \sum_{t=1}^{T} \frac{1}{N} \left(\sum_{i=1}^{N} \mathbb{E}\left[v_{i}(t)\right]\right)
\end{align}

We define the expected age of user $i$ at time $t$ as $\Delta_{i}^{\bm{p},\sigma}(t)$, i.e.,  $\Delta_{i}^{\bm{p},\sigma}(t)=E[v_{i}(t)]$. The overall age for user $i$ is 
\begin{align}\label{eq:2}
\Delta_i^{\bm{p},\sigma}=\frac{1}{T} \sum_{t=1}^T \Delta_i^{\bm{p},\sigma}(t)
\end{align}

For a given scheduling algorithm $\bm{p}$, the adversary aims to maximize $\Delta^{\bm{p},\sigma}$. We define a set $B(\bm{p})$, which consists of the solutions to the following optimization problem,
\begin{align}
    B(\bm{p}) = \argmax_{\sigma\in{\Sigma}} \Delta^{\bm{p},\sigma}
\end{align}
Similarly, for a given adversarial action $\sigma$, the base station aims to minimize $\Delta^{\bm{p},\sigma}$. We define a set $B(\sigma)$, which consists of the solutions to the following optimization problem,
\begin{align}
    B(\sigma) = \argmin_{\bm{p}\in{\mathcal{F}}} \Delta^{\bm{p},\sigma}
\end{align}
where $\mathcal{F}$ is the set of feasible probability distributions. Here, $B(\bm{p})$ and $B(\sigma)$ denote the best responses to actions $\bm{p}$ and $\sigma$.
 
A scheduling algorithm and an adversarial action, $(\bar{\bm{p}},\bar{\sigma})$ form a Nash equilibrium point if and only if $\Delta^{\bar{\bm{p}}, \bar{\sigma}} \leq \Delta^{\hat{\bm{p}},\bar{\sigma}}$ and $\Delta^{\bar{\bm{p}},\bar{\sigma}}\geq\Delta^{\bar{\bm{p}},\hat{\sigma}}$, for all $\hat{\bm{p}}\in{\mathcal{F}}$ and for all $\hat{\sigma}\in \Sigma$ \cite{han2019game}. In other words, $(\bar{\bm{p}},\bar{\sigma})$ is a Nash equilibrium point if and only if $\bar{\bm{p}}\in B(\bar{\sigma})$ and $\bar{\sigma}\in B(\bar{\bm{p}})$. 
 
We also provide a Stackelberg equilibrium point for this system model when the scheduling algorithm of the base station acts as a leader. In this two-step game, in the first step, for a fixed scheduling algorithm $\bm{p}$, the adversary aims to maximize the average age, thus, the adversary intends to apply an action from the set $B(\bm{p})$. In the second step, the base station selects the optimal scheduling algorithm to minimize the average age, i.e., chooses a scheduling algorithm $\bar{\bm{p}}$ which lies in the following set,
\begin{align}
     \bar{\bm{p}} \in \argmin_{\bm{p}\in\mathcal{F}} \Delta^{\bm{p},{\sigma}|{\sigma}\in{B(\bm{p})}}
\end{align}
We call $(\bar{\bm{p}},\bar{\sigma})$ a Stackelberg equilibrium point \cite{han2019game}, where $\bar{\sigma}\in{B(\bar{\bm{p}})}$. In Section~\ref{sec:NE}, we show that the Nash equilibrium for this setting does not exist, however, the Stackelberg equilibrium exists when base station acts as the leader.

\subsection{Communication Network Model with Diversity}\label{sub:2}
We consider a communication system, where at every time slot, the base station schedules a user from $N$ users following a stationary probability distribution, and chooses a sub-carrier from available $N_{sub}$ sub-carriers $(N_{sub}>1)$ following a stationary distribution to transmit update packets to the scheduled users. At a given time slot, the adversary present in the system can block a sub-carrier out of $N_{sub}$ sub-carriers, and in total it can block $\alpha T$ sub-carriers over the time horizon $T$. Similar to the previous setting, blocking a sub-carrier implies complete elimination of the update packet. As in the previous setting, the adversary only knows the scheduling algorithm used by the base station, and if the base station uses a randomized algorithm, then the adversary knows the probability distributions with which base station schedules the users and chooses the sub-carriers, but does not have access to the specific realizations of the actions taken by the base station. In the previous setting, we have a dedicated communication channel between a user and the base station, however, in the current setting, there is a pool of communication channels, namely, $N_{sub}$ communication channels, and at every time slot, the base station chooses one of these $N_{sub}$ communication channels with a valid probability mass function.

Let $\bm{p}$ be the probability mass function with which the base station schedules a user at every time slot, and let $\bm{q}$ be the probability mass function with which the base station chooses an sub-carrier at every time slot to transmit the update packet to the scheduled user. We denote the action of the adversary for the $j$th sub-carrier at time slot $t$ as $\sigma_{j}(t)$. Here $\sigma_{j}(t)=0$ means that the adversary blocks sub-carrier $j$ at time $t$, and $\sigma_{j}(t)=1$ means that the adversary does not block sub-carrier $j$ at time $t$. Thus, the action of the adversary to sub-carrier $j$ is a sequence of ones and zeros, and we denote this sequence as $\sigma_{j}$, and we call this sequence as the blocking sequence for sub-carrier $j$. We use $\sigma$ to denote the blocking matrix, whose $i$th row is the blocking sequence $\sigma_{j}$, for the $j$th sub-carrier, thus, the $(j,t)$th entry of $\sigma$, $\sigma_j(t)$ denotes the state of the communication channel on sub-carrier $j$ at time $t$. Thus, a feasible $\sigma$ should satisfy $\sum_{j=1}^{N_{sub}}\sum_{t=1}^{T} \sigma_j(t) \leq \alpha T$. We create a set $\Sigma$ of feasible $\sigma$.

From \cite{banarjee-infocom2022}, we know that, for this system model, the resulting age is finite, even when $T$ goes to infinity. Thus, we define the average age for this system model as
\begin{align} \label{objective}
\Delta^{\bm{p}, \bm{q}, \sigma} = \limsup_{T \to \infty} \frac{1}{T} \sum_{t=1}^{T} \frac{1}{N} \left(\sum_{i=1}^{N} \mathbb{E}\left[v_{i}(t)\right]\right)
\end{align}

For a given user scheduling algorithm $\bm{p}$ and a sub-carrier selection probability distribution $\bm{q}$, the adversary aims to maximize $\Delta^{\bm{p},\bm{q},\sigma}$. We define a set $B(\bm{p},\bm{q})$, which consists of the solutions to the following optimization problem
\begin{align}
    B(\bm{p},\bm{q}) = \argmax_{\sigma\in{\Sigma}} \Delta^{\bm{p},\bm{q},\sigma}
\end{align}
Similarly, for a given adversarial action $\sigma$, the base station aims to minimize $\Delta^{\bm{p}, \bm{q}, \sigma}$. We define a set $B(\sigma)$, which consists of the solutions to the following optimization problem
\begin{align}
    B(\sigma) = \argmin_{\bm{p}\in{\mathcal{F}}, \bm{q}\in{\mathcal{F}'}} \Delta^{\bm{p}, \bm{q},\sigma}
\end{align}
where $\mathcal{F}$ is the set of feasible probability distributions of choosing a user from $N$ users, and $\mathcal{F}'$ is the set of probability distributions of choosing a sub-carrier from $N_{sub}$ sub-carriers. Then $(\bar{\bm{p}},\bar{\bm{q}},\bar{\sigma)}$ is a Nash equilibrium point if and only if $(\bar{\bm{p}},\bar{\bm{q}})\in{B(\bar{\sigma})}$ and $\bar{\sigma}\in{B(\bar{\bm{p}},\bar{\bm{q}})}$. We show that Nash equilibrium for this setting exists, and find the Nash equilibrium.

\section{Nash Equilibrium For Model without Diversity}\label{sec:NE}
From \cite[Eqn.~(17)]{banarjee-infocom2022}, for a scheduling algorithm $\bm{p}$ and an adversarial action $\sigma$, the average age at time slot $(t+1)$ is 
\begin{align}\label{eq:8}
 \Delta_i^{\bm{p},\sigma}(t+1)= \sum_{\ell=1}^{t} \Gamma_i(\ell,t)+1
\end{align}
where $\Gamma_i(\ell,t)$ is defined in \cite[Eqn.~(15)]{banarjee-infocom2022}, i.e., 
\begin{align} \label{eq:9}
  \Gamma_i(k,\ell)=\prod_{j=k}^\ell \left(1-{\sigma}_{i}(j) p_{i}\right)
\end{align}

Without loss of generality, let us assume that the adversary blocks the middle $\alpha T$ slots consecutively for the first user. Without loss of generality, we assume that $(1-\alpha)T$ is a even number, thus $x_{1} = \frac{T} {2} - \frac{\alpha T}{2}$ is an integer. In addition, $T$ is large enough such that for a given $\alpha$, $(1-\alpha)T$ is also a large integer. Thus, $x_{1}$ is large. From (\ref{eq:8}) and (\ref{eq:9}), for $1<j\leq N$,
\begin{align}\label{eq:10}
    \hspace*{-0.2cm} \Delta_{j}^{\bm{p},\sigma}= &\frac{1}{T}\Big[(T-1) (1-p_{j}) + (T-2) (1-p_{j})^{2} + \cdots + \nonumber\\
    & \quad (T -x_{1}) (1-p_{j})^{x_{1}} + \cdots + (1-p_{j})^{T-1} + T \Big] \hspace*{-0.1cm} \\ 
    = &\frac{1}{T}\Big[T \big((1-p_{j}) + (1-p_{j})^{2} + \cdots \nonumber\\
    & \quad + (1-p_{j})^{x_{1}}+\cdots+(1-p_{j})^{T}\big)\nonumber \\
    & \quad -\big((1-p_{j}) +  2 (1-p_{j})^{2} + \cdots \nonumber\\
    & \quad + x_{1} (1-p_{j})^{x_{1}} + \cdots + T (1- p_{j})^{T}\big) + T\Big]
\end{align}
For large enough $T$, (\ref{eq:10}) can be approximated as,
\begin{align}\label{eq:11}
    \Delta_{j}^{\bm{p},\sigma} =& \frac{1}{T} \Big[T \frac{1-p_{j}}{p_{j}} - \frac{(1-p_{j})}{p_{j}^{2}} + T\Big]\\
    = & \frac{1}{T}\Big[\frac{T}{p_{j}}  - \frac{1}{p_{j}^{2}}  + \frac{1}{p_{j}}\Big]
\end{align}
Using our assumption $T\gg\frac{1}{p_{j}}$, (\ref{eq:11}) can be approximated as 
\begin{align}\label{eq:12}
     \Delta_{j}^{\bm{p},\sigma} = \frac{1}{p_{j}} 
\end{align}
Similarly, from (\ref{eq:8}) and (\ref{eq:9}), the average age of the first user,  
\begin{align}\label{eq:13}
    \Delta_{1}^{\bm{p},\sigma} = & \frac{1}{T}\Big[2\big((1+\alpha T) (1- p_{1})^{x_{1}} + (2+\alpha T) (1-p_{1})^{x_{1}-1} \nonumber \\
    & \quad +\cdots +(x_{1} + \alpha T) (1-p_{1})\big) + (1-p_{1})^{2} \nonumber\\
    & \quad +2 (1-p_{1})^{3} + \cdots +x_{1} (1-p_{1})^{x_{1}+1} + \nonumber \\ 
    & \quad (x_{1}-1) (1-p_{1})^{x_{1}+2} + (1-p_{1})^{2 x_{1}} \nonumber\\
    & \quad + \frac{\alpha T (1+\alpha T)}{2}+T\Big]
\end{align}
Note that $x_{1}+\alpha T = \frac{T} {2} + \frac{\alpha T} {2}$. For large $T$, with similar manipulations used to get (\ref{eq:12}), (\ref{eq:13}) can be approximated as,
\begin{align}\label{eq:cont_eq}
    \Delta_{1}^{\bm{p},\sigma}  = (1+\alpha) \frac{1-p_{1}} {p_{1}} + \frac{\alpha (1+ \alpha T)}{2}+1
\end{align}

Thus, summing (\ref{eq:12}) over all $j$, and (\ref{eq:13}), the total average age for all $N$ users is
\begin{align}\label{eq:15}
    \Delta^{\bm{p},\sigma} = \frac{1}{N} \left( \sum_{j=2}^N \frac{1}{p_j} \!+\! (1\!+\!\alpha) \frac{1\!-\!p_{1}} {p_{1}} \!+\! \frac{\alpha (1\!+\! \alpha T)}{2} \!+\!1 \right)
\end{align}
In the rest of this paper, the base station optimizes over $\bm{p}$ and the adversary optimized over $\sigma$, thus removing all the constants from (\ref{eq:15}),  $\Delta^{\bm{p},\sigma}_{eq}$, 
\begin{align}\label{eq:20}
    \Delta^{\bm{p},\sigma}_{eq} = \sum_{j=2}^{N} \frac{1}{p_{j}} +  \frac{(1+\alpha)}{p_{1}} - \alpha + \frac{\alpha (1 + \alpha T)} {2} 
\end{align}

In the following theorem, we find the probability distribution $\bm{p}$ which minimizes (\ref{eq:15}).

\begin{theorem}\label{th:1}
For large enough $T$, if the adversary blocks user $1$ for $\alpha T$ time slots consecutively in the middle of the time horizon, then the optimal choice for the base station is to schedule the users with the following probability distribution, $p_1 = \frac{\sqrt{1+\alpha}}{N-1+\sqrt{1+\alpha}}$, and
$p_j = \frac{1}{N-1+\sqrt{1+\alpha}}$, for $j \neq 1$.
\end{theorem}

\begin{Proof}
Using the objective function in (\ref{eq:20}) together with the constraint that $\sum_{j=1}^{N} p_j=1$ yields the Lagrangian as
\begin{align}
\mathcal{L} =  \sum_{j=2}^N \frac{1}{p_j} + \frac{(1+\alpha)}{p_1} +\lambda \left(\sum_{j=1}^{N} p_j-1\right)
\end{align}
The KKT conditions are
\begin{align}
\frac{1+\alpha}{p_1^2} = \lambda, \quad \frac{1}{p_j^2} = \lambda, \ j \neq 1
\end{align}
Using $\sum_{j=1}^{N} p_j=1$ to find $\lambda$ yields the optimum solution as
\begin{align}
p_1 = \frac{\sqrt{1+\alpha}}{N\!-\!1\!+\!\sqrt{1+\alpha}}, \quad 
p_j = \frac{1}{N\!-\!1\!+\!\sqrt{1+\alpha}}, \ j \neq 1
\end{align}
completing the proof.
\end{Proof}

In the following theorem, we prove the converse for Theorem~\ref{th:1}. 

\begin{theorem}\label{th:2}
If the base station schedules the users with a probability distribution that satisfies $p_{1}\geq p_{2}\geq \cdots\geq p_{N}$, then an optimal action for the adversary is to block user $N$ consecutively in the middle of time horizon for $\alpha$ T time slots.
\end{theorem}

\begin{Proof}
Consider that the adversary blocks the users for $\alpha_{1}T, \alpha_{2} T, \cdots,\alpha_{N} T$ time slots over the time horizon, where $\sum_{j=1}^{N} \alpha_j = \alpha$, and we denote this blocking action of the adversary as $\bar{\sigma}$. Thus, the average age of this system is ${\Delta}^{\bm{p},\bar{\sigma}}$. Note that the blocking sequences of $\bar{\sigma}$ are arbitrary, i.e., they need not be consecutive or in the middle of the time horizon. Now, we assume that the adversary blocks the users for $\alpha_{1}T , \alpha_{2}T, \cdots, \alpha_{N}T$ time slots consecutively in the middle of the time horizon, and we denote this blocking action of the adversary as $\tilde{\sigma}$. It is obvious that this adversary violates the constraint that at a time slot it can block only $1$ user. Let us denote the average age of this system as ${\Delta}^{\bm{p},\tilde{\sigma}}$. From \cite[Thm.~5]{banarjee-infocom2022}, we know that ${\Delta}^{\bm{p},\tilde{\sigma}} > {\Delta}^{\bm{p},\bar{\sigma}}$. For large $T$, similar to (\ref{eq:cont_eq}), we can approximate ${\Delta}_{i}^{\bm{p},\tilde{\sigma}}$ as $\tilde{\Delta}_{i}^{\bm{p},\tilde{\sigma}} = (1 + \alpha_{i}) \frac{1- p_{i}}{p_{i}} + \frac{\alpha_{i} (1+\alpha_{i} T)}{2}+1$. Noting that $\sum_{i=1}^N \alpha_i$ is constant and $\bm{p}$ is given and constant, we have,
\begin{align}
   \!\! \tilde{\Delta}^{\bm{p},\tilde{\sigma}}_{eq}
    = & \sum_{i=1}^{N} \frac{1}{p_{i}}  - \sum_{i=1}^{N} \alpha_{i}  + \sum_{i=1} ^{N} \frac{\alpha_{i} } {p_{i}} + \sum_{i=1}^{N} \frac{\alpha_{i} }{2} + \sum_{i=1}^{N} \frac{\alpha_{i}^{2} T}{2} \\
    = & C + \sum_{i=1}^{N} \frac{\alpha_{i}}{p_{i}} + \sum_{i=1}^{N} \frac{\alpha_{i}^{2} T } {2}
\end{align}
where $C$ is the appropriate constant. Now, if we pick any one of the $\alpha_{i}=\alpha$ and the rest of the $\alpha_{i}=0$, then this selection maximizes $\sum_{i=1}^{N} \frac{\alpha_{i}^{2}}{2}$. Let us define $k= \argmax_{i}{\frac{1}{p_{i}}}$. If the adversary chooses $\alpha_{k} = \alpha$ and the rest of the $\alpha_{i} = 0$, then this selection maximizes $\sum_{i=1}^{N} \frac{\alpha_{i}}{p_{i}}$. Let us denote this particular choice of the adversarial action as $\hat{\sigma}$. Note that $\hat{\sigma}$ is a feasible adversarial action. Let the average age of the system corresponding to $\hat{\sigma}$ be ${\Delta}^{\bm{p},\hat{\sigma}}$. Thus, ${\Delta}^{\bm{p},\hat{\sigma}}\geq{\Delta}^{\bm{p},\tilde{\sigma}}\geq{\Delta}^{\bm{p},\bar{\sigma}}$, which concludes the proof.
\end{Proof}

In the next theorem, we show that for this setting the Nash equilibrium does not exist.

\begin{theorem}
If a base station schedules $1$ user out of $N$ users at every time slot following a stationary distribution, and if an adversary can block $1$ out of $N$ communication channels between the base station and the users with a constraint that it can block maximum $\alpha T$ communication channels, and if the objective of the base station is to minimize the average age, and the objective of the adversary is to maximize the average age, then a Nash equilibrium for this problem does not exist.    
\end{theorem}

\begin{Proof}
We prove this theorem by contradiction. Let us assume that $(\bm{p}',\sigma')$ is a Nash equilibrium point for this problem, where $\bm{p}'$ is a valid probability distribution and $\sigma'$ is a valid blocking matrix. Then according to Theorem~\ref{th:2}, the adversary needs to block only the user which has the lowest probability of getting scheduled by the base station, for $\alpha T$ time slots in the middle of the time horizon, otherwise $\sigma'\not\in{B(\bm{p}')}$. Without loss of generality, assume that $p_{1}'\geq p_{2}'\geq\cdots \geq p_{N}'$, and the adversary blocks only the user $N$ for $\alpha T$ time slots in the middle of the time horizon. From Theorem!\ref{th:1}, we know that the optimal scheduling algorithm for the base station is to schedule user $N$ with probability $\frac{\sqrt{1+\alpha}}{N-1+\sqrt{1+\alpha}}$ and schedule other users with probability $\frac{1}{N-1+\sqrt{1+\alpha}}$, thus user $N$ gets scheduled by the base station with the highest probability, which contradicts our assumption that $p_{1}'\geq p_{2}'\geq\cdots \geq p_{N}'$. 
\end{Proof}

\section{Stackelberg Equilibrium For Model without Diversity}
Let $\bm{p}^{*}$ be an optimal solution for the following optimization problem $\argmin_{\bm{p}\in{\mathcal{F}}} \Delta^{\bm{p}, B(\bm{p})}$. Recall that, we say the pair $(\bm{p}^{*}, \sigma^{*})$ is a Stackelberg equilibrium pair, if $\sigma^{*}\in{B(\bm{p}^{*})}$. 

In the following theorem we find a Stackelberg equilibrium point for the system model with no diversity.

\begin{theorem}
If the base station schedules the users with uniform probability distribution and if the adversary blocks any one of the $N$ communication channels consecutively in the middle of the time horizon for $\alpha T$ time slots, then the pair of these policies is a Stackelberg equilibrium point for the communication network setting with no diversity.
\end{theorem}

\begin{Proof}
From Theorem~\ref{th:2}, we know that if the base station schedules the users with a probability distribution $\bm{p}$ where $p_{1}\geq p_{2} \geq \cdots \geq p_{N}$, then the best action for the adversary is to block the $N$th user consecutively in the middle of time horizon for $\alpha T$ time slots. For large enough $T$, from (\ref{eq:20}), the average age of the system for this pair of actions is 
\begin{align}
     \Delta^{\bm{p},\sigma}_{eq} & =   \sum_{j=1}^{N-1} \frac{1}{p_{j}} +  \frac{(1+\alpha)}{p_{N}} - \alpha + \frac{\alpha (1 + \alpha T)} {2} \\ & = C + \sum_{i=1}^{N-1} \frac{1} {p_{i}} + \frac{(1 + \alpha)}{p_{N}}
\end{align} 
where $C$ is a constant. Thus, the optimization problem is
\begin{align}
\argmin_{\bm{p}} & \quad \sum_{i=1}^{N-1} \frac{1} {p_{i}} + \frac{(1 + \alpha)}{p_{N}} \nonumber\\
\textrm{s.t.} & \quad \sum_{i=1}^{N}p_{i} =1, \quad \text{and}
\quad p_{1} \geq p_{2} \geq\cdots\geq p_{N}
\end{align}
The Lagrangian for this problem is
\begin{align}
    \!\!\!\!\mathcal{L} \!=\! \sum_{i=1}^{N-1} \! \frac{1} {p_{i}} \!+\! \frac{(1 \!+\! \alpha)}{p_{N}} \!+\! \lambda \Big(\sum_{i=1}^{N}p_{i}\!-\!1\Big) \!+\! \sum_{i=1}^{N-1} \mu_{i} (p_{i+1}\!-\!p_{i}) \!\!
\end{align}
The KKT conditions are
\begin{align}
&p_1 = \frac{1}{\sqrt{\lambda-\mu_1}}, \qquad \quad
p_N = \frac{\sqrt{1+\alpha}}{\sqrt{\lambda+\mu_{N-1}}} \\
&p_i = \frac{1}{\sqrt{\lambda-\mu_{i-1}-\mu_i}}, \quad i=2,\ldots,N-1
\end{align}
Now, if $\mu_{N-1}=0$, then $p_{N}= \frac{\sqrt{1+\alpha}}{\sqrt{\lambda}}$ and $p_{N-1}= \frac{1}{\sqrt{\lambda + \mu_{N-2}}}$. Because of $\mu_{N-2}\geq 0$ and $\alpha> 0$, this implies $p_{N-1}< p_{N}$, which violates the primal feasibility. Thus, $\mu_{N-1}>0$. With similar arguments, $\mu_{i}>0$, for all $i=1,\cdots,N$. Now, from complementary slackness, we have $p_{1}=p_{2}=\cdots=p_{N}$.  
\end{Proof}

\section{Nash Equilibrium For Model with Diversity }
First, we find the optimal strategy for the base station for a specific blocking strategy for the adversary.

\begin{theorem}\label{th:5}
If the adversary blocks the sub-carriers uniformly and consecutively for $\alpha T$ time slots in the middle of the time horizon, then the optimal strategy for the base station is to schedule the users with uniform distribution and choose the sub-carriers with uniform distribution.
\end{theorem}

\begin{Proof}
Let us assume that the base station schedules a user among $N$ users with the probability distribution $\bm{p}$ and chooses a sub-carrier among $N_{sub}$ sub-carriers with the probability distribution $\bm{q}$. Consider that the adversary is blocking a sub-carrier among $N_{sub}$ sub-carriers only for $\alpha T$ time slots with uniform distribution. Let this action of the adversary be $\sigma'$. The age in time slot $t+1$ where $\frac{T}{2} - \frac{\alpha T}{2}< t \leq\frac{T}{2} + \frac{\alpha T}{2}$ is
\begin{align}\label{eq:28}
    &\Delta_{i}^{\bm{p}, \bm{q}, \sigma'}(t+1) \nonumber\\
    = &(\Delta_{i}^{{\bm{p}, \bm{q}, \sigma'}}(t)+1)  (1-p_{i})\nonumber + p_{i}  (\Delta_{i}^{\bm{p}, \bm{q}, \sigma'}(t)+1)  \sum_{j=1}^{N_{sub}} \frac{q_{j}}{N_{sub}}  \nonumber \\ 
    & \qquad + p_{i}  \sum_{j=1}^{N_{sub}} q_{j} \left(1-\frac{1}{N_{sub}}\right)  \\ 
    = & (\Delta_{i}^{\bm{p}, \bm{q}, \sigma'}(t)+1) (1-p_{i}) + \frac{p_{i} (\Delta_{i}^{\bm{p}, \bm{q}, \sigma'}(t)+1)}{N_{sub}} \nonumber \\ 
    & \qquad + p_{i}\left(1-\frac{1}{N_{sub}}\right) \\
    = & (\Delta_{i}^{\bm{p}, \bm{q}, \sigma'}(t)+1) \left(1 - p_{i}\left(1-\frac{1}{N_{sub}}\right)\right) +p_{i}\left(1-\frac{1}{N_{sub}}\right) \\
    = &  \Delta_{i}^{\bm{p}, \bm{q}, \sigma'}(t)\left(1-p_{i}\left(1-\frac{1}{N_{sub}}\right)\right)+1
\end{align}
From \cite{banarjee-infocom2022}, for $t\leq \frac{T}{2} -\frac{\alpha T}{2}$ and for $t> \frac{T}{2} + \frac{\alpha T}{2}$, the average age for time slot $t+1$ is 
\begin{align}
    \Delta_{i}^{\bm{p}, \bm{q}, \sigma'}(t+1) = \Delta_{i}^{\bm{p}, \bm{q}, \sigma'}(t) (1-p_{i}) + 1
\end{align}

In the previous system model, i.e., communication network with no diversity, the adversary completely blocked one user in the middle of the time horizon for $\alpha T$ time slots. However, in this system model, the probability of getting scheduled by the base station for user $i$ is reduced to $p_{i}\left(1-\frac{1}{N_{sub}}\right)$ from $p_{i}$, for $\alpha T$ time slots in the middle of the time horizon $T$. Using (\ref{eq:8}) and (\ref{eq:9}) and using similar approximations to those in obtaining (\ref{eq:cont_eq}) from (\ref{eq:13}), we get the average age for user $i$ as
\begin{align}\label{eq:31}
    \!\!\!\!\Delta_{i}^{\bm{p}, \bm{q}, \sigma'} =&  \frac{2 x_{1} (1-p_{i})}{p_{i}} + \frac{\alpha T \left(1- p_{i} \left(1-\frac{1}{N_{sub}}\right)\right) } {p_{i} \left(1-\frac{1}{N_{sub}}\right)} + T \\
    =&  \frac{2 x_{1}}{p_{i}} + \frac{\alpha T} {p_{i} \left(1-\frac{1}{N_{sub}}\right)}
\end{align}
where $x_{1} = \frac{T}{2} - \frac{\alpha T}{2}$. Note that, for each user, this system can be thought of as three blocks with the absence of the adversary in each block. First block is from time slot $1$ to time slot $\frac{T}{2} - \frac{\alpha T}{2}$ with probability of getting scheduled by the base station $p_{i}$. Second block is from time slot $\frac{T}{2} - \frac{\alpha T}{2} + 1$ to time slot $\frac{T}{2} + \frac{\alpha T}{2}$ with probability of getting scheduled by the base station $p_{i}\left(1-\frac{1}{N_{sub}}\right)$. The third block is from time slot $\frac{T}{2} + \frac{\alpha T}{2} +1$ to time slot $T$ with probability of getting scheduled by the base station $p_{i}$. Thus, we get 
\begin{align} \label{eq:39}
    \Delta^{\bm{p}, \bm{q} , \sigma'} =& \sum_{i=1}^{N} \frac{1-\alpha}{p_{i}} + \frac{\alpha} {p_{i}\left(1 - \frac{1}{N_{sub}}\right)} = \sum_{i=1}^{N} \frac{C_{1}}{p_{i}} 
\end{align}
where $C_{1}$ is a constant. Minimizing (\ref{eq:39}) with respect to $\bm{p}$ subject to $\sum_{j=1}^N p_j=1$, we obtain $p_{i}=\frac{1}{N}$. As (\ref{eq:39}) does not depend on $\bm{q}$, we can choose any valid probability distribution, and we choose the uniform distribution.
\end{Proof}

The next theorem gives the optimal blocking sequence for the adversary when the base station chooses the users and the sub-carriers uniformly.

\begin{theorem}
The triplet $(\bar{\bm{p}}, \bar{\bm{q}}, \sigma')$ is a Nash equilibrium for the model with diversity, where $\bar{\bm{p}}$, $\bar{\bm{q}}$ are uniform distributions over $N$ users and $N_{sub}$ sub-carriers, respectively, and $\sigma'$ blocks the sub-carriers uniformly and consecutively for $\alpha T$ time slots in the middle of the time horizon.
\end{theorem}

\begin{Proof}
From \cite[Thm.~4]{banarjee-infocom2022}, when the scheduling algorithm chooses the users and the sub-carriers uniformly, the optimal strategy for the adversary is to block the sub-carriers consecutively and in the middle of the time horizon. Thus, $\sigma' \in B(\bar{\bm{p}},\bar{\bm{q}})$. From Theorem~\ref{th:5} in this paper,  $(\bar{\bm{p}},\bar{\bm{q}})\in B(\sigma')$. Thus, $(\bar{\bm{p}}, \bar{\bm{q}}, \sigma')$ is a Nash equilibrium point. 
\end{Proof}

\bibliographystyle{unsrt} 
\bibliography{references}

 \end{document}